\documentclass{raa}        
\usepackage{graphicx,times}
\usepackage{natbib}
\usepackage{amssymb,amsmath}
\bibpunct{(}{)}{;}{a}{}{,}

\usepackage[a4paper=true,dvipdfm=true,pagebackref=true]{hyperref}
\hypersetup{pdftitle = The title of my PDF, pdfauthor = My name, pdfsubject= The subject, pdfkeywords = keyword1 keyword2 keyword3} 
\hypersetup{colorlinks = true, linkcolor = green, anchorcolor = red, citecolor = blue, filecolor = red, pagecolor = red, urlcolor = red}

\begin{document}

   \title{The observed temperature distribution of contact binaries derived from SDSS photometry}

 \volnopage{ {\bf 2019} Vol.\ {\bf X} No. {\bf XX}, 000--000}
   \setcounter{page}{1}

   \author{Alexander Kurtenkov\inst{}, Mirela Napetova\inst{}
   }

   \institute{ Institute of Astronomy and National Astronomical Observatory, Bulgarian Academy of Sciences, 72 Tsarigradsko Shose Blvd., 1784 Sofia, Bulgaria; {\it al.kurtenkov@gmail.com}\\
\vs \no
   {\small Received 2019 February 14; accepted 20XX Month Day}
}

\abstract{Data from the VSX, SDSS DR12, and Gaia DR2 catalogs were combined in order to explore the observed temperature distribution of known contact binaries. Color-temperature relations were applied to calculate the effective temperatures of approximately 20000 binaries and a temperature distribution was built using the results with a precision better than 160 K. The temperature distribution shifts to higher temperatures when increasing the distance of the observed objects. The distribution contains a a previously detected bump in the 6000-7000 K range. It is shown that this second maximum is caused by distant and highly reddened objects that are systematically more luminous than contact binaries in this temperature range. We argue that these objects are other variables, possibly semi-detached binaries, misclassified as contact binaries and that the actual temperature distribution of contact binaries does not contain such a peak.}


   \authorrunning{A. Kurtenkov \& M. Napetova}            
   \titlerunning{The observed temperature distribution of contact binaries derived from SDSS photometry}  
   \maketitle

%
\section{Introduction}           
\label{sect:intro}
Binary stars have for long been the source of our knowledge of fundamental stellar parameters and our way of testing stellar evolutionary models. Binaries of the W UMa (or EW) type were first recognized because of their lightcurve morphology \citep{2009ebs..book.....K}. These contact binaries show no phase intervals of constant brightness, making it hard to distinguish between eclipses due to the highly deformed shapes of the components overfilling their Roche lobe surfaces \citep{1970VA.....12..217B}. Studies show that a W UMa binary system usually consists of two main-sequence stars of spectral types F, G, or K. The components share a common atmosphere and due to energy transport their effective temperatures are nearly equal \citep{2005ApJ...629.1055Y}. 

The major goal of this work is to derive and interpret the temperature distribution of contact binaries as observed at the current stage of our technical progress. Evolutionary factors are expected to play a main role in the W UMa temperature distribution. According to \citet{1998AJ....116.2998R} contact binaries can be a stage in an angular-momentum-loss-driven evolution of a merge forming only one single star. \citet{1998AJ....116.2998R} also pointed out that their spread is an indication of their advanced age as they are not to be found in young open clusters. Therefore W UMa binaries with large surface temperatures are not observed. Another major factor is that the initial mass function (IMF) is decreasing for the majority of the stellar mass range \citep{1955ApJ...121..161S}. As temperature increases with mass along the main sequence, the IMF shape should lead to a decrease in the temperature distribution as well. On the other hand the distribution may decline towards lower temperatures since low-mass stars are less probable to be a part of binary or multiple systems \citet{2013ARA&A..51..269D}. These and other factors form the actual temperature distribution of contact binaries. It is then influenced by a number of selection effects to form the observed distribution. Most importantly, cooler objects along the main sequence are also smaller and much less luminous. Hence, hotter contact binaries are more easily discovered, which should cause the low end of the observed temperature distribution to sink.

The observed temperature distribution can be based either on spectroscopic observations or on photometric colors. In the current work we aim to do the latter. The number of known contact binaries has been increasing rapidly in the past decade thanks to large area variability surveys such as Kepler \citep{2011AJ....141...83P} and Catalina \citep{2014ApJS..213....9D}. The higher statistical sample is therefore a major advantage of using photometric data for this purpose. Nevertheless, photometric temperature estimates are less reliable for individual objects as they are influenced by interstellar reddening. A recent spectroscopy-based temperature distribution of W UMa stars was obtained by \citet{2017RAA....17...87Q}. A comparison to our results is made and discussed in this paper.

\section{Calculating temperatures of contact binaries based on SDSS colors}           
\label{sect:method}
Three large catalogs are utilized for our task: the VSX, SDSS DR12 and Gaia DR2. In the age of large sky surveys the most comprehensive source of catalog data on variable stars is the International variable star index (VSX) of the American Association of Variable Star Observers (AAVSO) \citep{2006SASS...25...47W}. The Sloan Digital Sky Survey (SDSS) Legacy Survey contains well-calibrated astrometry and optical (\textit{ugriz}) photometry for 230 million celestial objects over 740 square degrees on the sky \citep{2012ApJS..203...21A}. The second \textit{Gaia} data release provides extremely precise astrometry, complemented by photometry, and some astrophysical parameters for sources brighter than 21 mag (1.7 billion sources) \citep{2018A&A...616A...1G}. Gaia parallaxes and proper motions are available for 1.3 billion sources.

As of January 2019 a query for all EW variables in VSX would return 70465 records, with only 77 missing period data and 73 missing photometric amplitude data. We matched these to the SDSS DR12 and Gaia DR2 data using search radii of 2$''$ and 1.5$''$ respectively. Thus we combined the available periods, amplitudes and Sloan $griz$ magnitudes for a sample of 20085 objects classified as W UMa-type binaries, including Gaia parallaxes for 19930 of them. Effective temperature estimates and G-band extinction estimates from the Gaia Apsis pipeline are available for 76\% and 51\% of the objects respectively.

We chose the SDSS catalog to calculate color temperatures for two main reasons:
\begin{itemize}
  \item \textbf{It is deep.} The $g$-band photometric error is less than 0.02\,mag for 99.3\% of our sample, whereas for 98.9\% of the sample $g<20$\,mag. This ensures that the propagation of photometric errors into temperatures is negligible.
  \item \textbf{The frames in different passbands are taken almost simultaneously.} Actually, the time cadence between two filters is 71.72\,s as the order of imaging is $riuzg$. The low cutoff in the W UMa period distribution is at $\sim0.22$\,d. Even for these short-period contact binaries the time interval between the $r$ and $g$ images of $\sim4.8$\,min translates to a phase difference of just 0.015, which is negligible in terms of brightness change. This is the reason we preferred SDSS to the more complete Pan-STARRS1 catalog \citep{2016arXiv161205560C}.   
\end{itemize}

Due to the temperature difference of the components and the gravity darkening effect the colors of contact binaries vary slightly with phase. As the magnitude of both effects is $\sim10^{2}$\,K, a satisfactory estimate of the temperature of the common atmosphere can be made using multicolor photometry at a given phase. To obtain the temperatures from the SDSS colors we applied the color-temperature relations derived by \citet{2013ApJ...771...40B}. They are of the form
\begin{eqnarray}
    T_{X-Y} &=& c_{0}+c_{1}(X-Y)+c_{2}(X-Y)^{2}+c_{3}(X-Y)^{3},                           
\end{eqnarray}
where $(X-Y)$ represents the color index and $T_{X-Y}$ is the effective photospheric temperature. The coefficients derived by Boyajian et al. (2013) for the Sloan photometric system are given in Tab.\,\ref{color-T}.

\begin{table}[!ht]
\begin{center}
\caption[Polynomial coefficients of the color-temperature relations from \citep{2013ApJ...771...40B}.]{Polynomial coefficients of the color-temperature relations from \citep{2013ApJ...771...40B}.}\label{color-T}
\small
\begin{tabular}{c@{ }c@{ }c@{ }c@{ }c@{ }c@{ }}
\hline\hline
 color \,& $c_{0}$  & $c_{1}$ & $c_{2}$ & $c_{3}$ & range of applicability [mag] \\
\hline
 $g-r$	\,\,& 7526	\,\,& -5570   \,\,& 3750 \,\,&  -1332.9	\,\,& (-0.23,1.40)  \\
 $g-i$	\,\,& 7279	\,\,& -3356   \,\,& 1112 \,\,&  -153.9	\,\,& (-0.43,2.78)  \\
 $g-z$	\,\,& 7089	\,\,& -2760   \,\,& 804  \,\,&  -95.2	\,\,& (-0.58,3.44)  \\
\hline        
\end{tabular} 
\end{center}  
\end{table}  

Thus we obtained three color temperature values for each object: $T_{g-r}$, $T_{g-i}$ and $T_{g-z}$. We adopted the median of the three as our final result for the photospheric temperature and the rms of the three as the error.

\section{Results and discussion}
\label{sect:results}
\subsection{The observed temperature distribution}

\begin{figure}[!h]
\centering
\includegraphics[width=10cm]{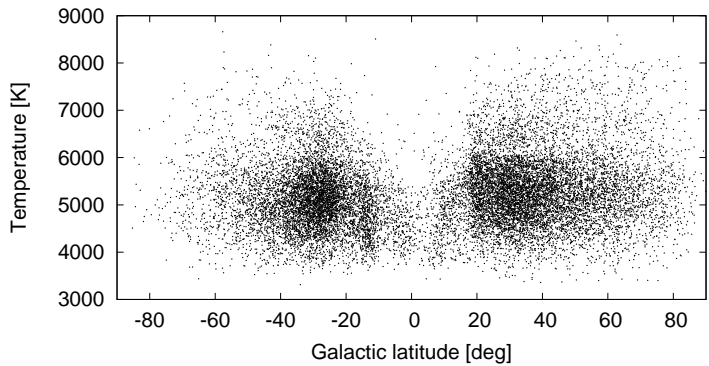}
\caption{Galactic latitude - Temperature diagram of the SDSS based data for contact binaries. The few objects close to the galactic plane are systematically redder.}
\label{gallat}
\end{figure}

To build the observed distribution we filtered the data in the following manner:
\begin{itemize}
  \item The objects with colors outside the applicability range were excluded. All three applicability ranges cover $3500-8900$\,K. As it is later shown, this is not a large restriction, since the vast majority of known contact binaries lie within this range. In order to avoid any systematic effects from this restriction, we will only analyse the temperature distribution in the $4000-8000$\,K range.
  \item The objects with rms larger than 160\,K were excluded. These are arbitrarily considered as successful temperature estimates and make up $\sim$11\% of the sample.
  \item The objects at galactic longitudes $|b|<10^{\circ}$ are excluded in order to mitigate contamination by strong reddening (Fig.\,\ref{gallat}).
\end{itemize}

\begin{figure}[!h]
\centering
\includegraphics[width=12cm]{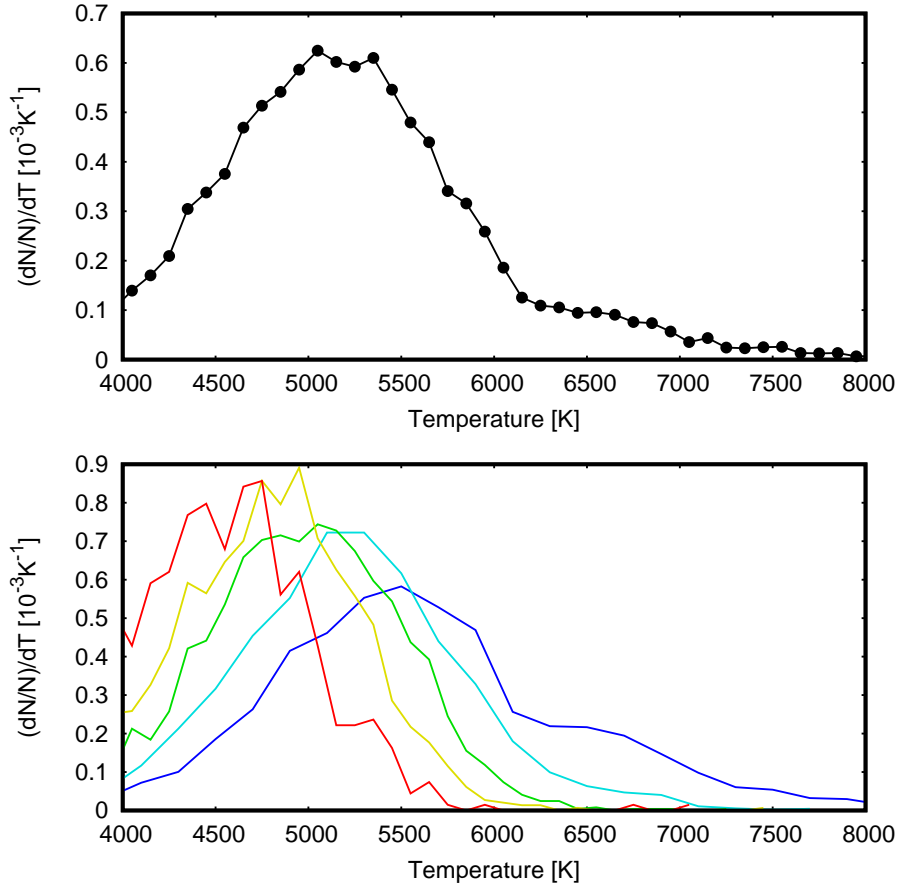}
\caption{\textit{Above}: The derived temperature distribution of contact binaries over the $4000-8000$\,K range. \textit{Below}: The temperature distributions of contact binaries in 5 Gaia parallax ranges, from blue to red (right to left): $0.1-0.4$\,mas, $0.4-0.7$\,mas, $0.7-1.0$\,mas, $1.0-1.5$\,mas, and $>1.5$\,mas. No dereddening procedures are applied on either diagram.}
\label{tempdist}
\end{figure}

The results are presented on Fig.\,\ref{tempdist}. The distribution has a somewhat flat peak at $\sim5200$\,K, and although this value does not have a physical meaning, it can be used for comparison. A broad blue bump is present at $\sim6500$\,K.

We demonstrate the aforementioned selection bias towards the brighter hotter objects by dividing the data into five groups of similar size at different distances, within the Gaia parallax ranges of $0.1-0.4$\,mas, $0.4-0.7$\,mas, $0.7-1.0$\,mas, $1.0-1.5$\,mas, and $>1.5$\,mas respectively. The five temperature distributions are visibly shifting to the right with distance as they also become somewhat flatter (Fig.\,\ref{tempdist}). The approximate peaks are at $4600$\,K, $4800$\,K, $5000$\,K, $5200$\,K, and $5500$\,K from closest to farthest.

\subsection{Data comparison}

\citet{2017RAA....17...87Q} published a catalog of W UMa binaries as observed by LAMOST -- Large Sky Area Multiobject Fibre Spectroscopic Telescope \citep{1996ApOpt..35.5155W,2012RAA....12.1243L,2015RAA....15.1095L}. They list 5363 binaries that have spectral parameters fitted (temperature, $\log g$, RV, metallicity). Their data shows a peak in the temperature distribution at 5700\,K with a blue bump present at 6600\,K and an overall shape very similar to the one we derived. Our results are visibly shifted to the right to those from LAMOST, which could be due to the lack of a reddening correction.

We compare the accuracy of our temperature to the estimates from LAMOST DR4 \citep{1996ApOpt..35.5155W,2012RAA....12.1243L,2015RAA....15.1095L} and from the Gaia Apsis pipeline \citep{2018A&A...616A...8A}. The statistics are based on 2021 matches with LAMOST and 10935 matches with Gaia. The temperatures derived from SDSS show a median offset from the Gaia temperature of -86 K and a standard deviation of 222 K. The median offset from the LAMOST estimates is -209 K with a standard deviation of 395 K. 

\begin{figure}[!h]
\centering
\includegraphics[width=10cm]{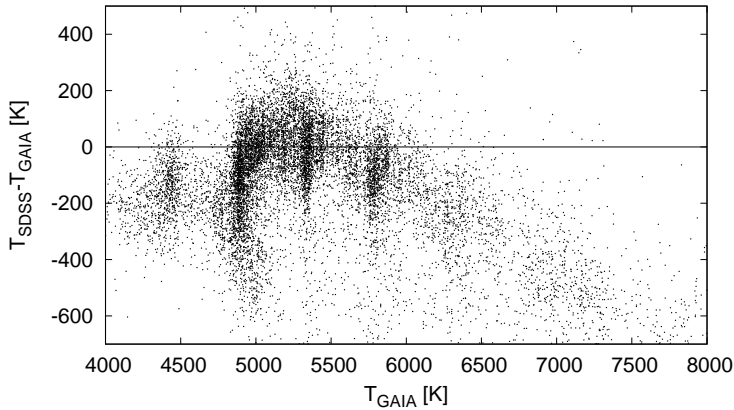}
\caption{Differences between Gaia and SDSS-derived temperatures plotted versus the Gaia temperatures. The vertical patterns indicate artificial discretization of the data.}
\label{gaiatemp}
\end{figure}

Visual inspection of temperature differences from the LAMOST results plotted versus colors shows no specific trends. The plots of Gaia temperatures versus temperature differences from SDSS show vertical striped patterns indicating an artificial discretization of the Apsis temperatures (Fig.\,\ref{gaiatemp}). The Gaia temperatures are obtained by running a machine learning algorithm on $G_{BP}-G$ and $G-G_{RP}$ color data. It uses temperatures from APOGEE, Kepler, LAMOST, RAVE and the RVS auxiliary catalog as training sets. These are not distributed uniformly, which could be the cause for the observed patterns. Especially dominant is the peak density at 4900\,K, which coincides with a peak in the Apsis training set.

\subsection{Explanation of the blue bump}
\citet{2017RAA....17...87Q} interpret the central peak and the blue bump as actual features corresponding to the p-p chain and CNO cycle nuclear reactions, but we do not see any evidence supporting this claim. The initial mass function is not known to contain similar bumps and there are no apparent reasons for stars, where the CNO-cycle is more efficient, to be more common than cooler ones.

\begin{figure}[!h]
\centering
\includegraphics[width=14cm]{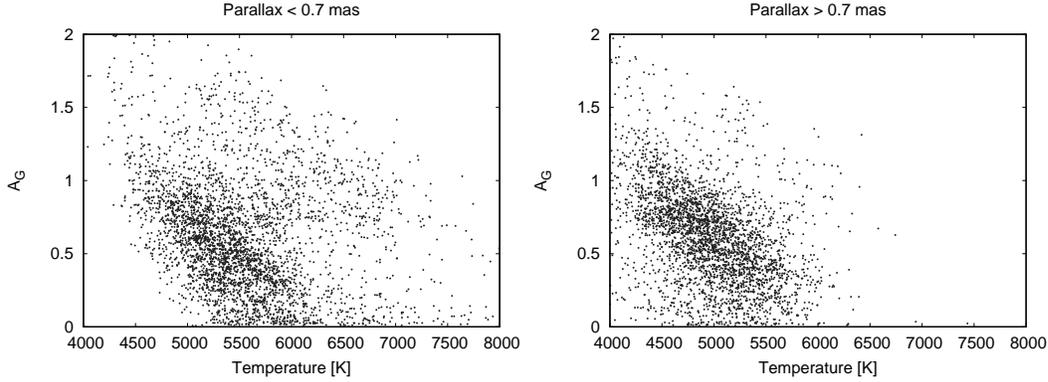}
\caption{Temperature-extinction diagrams for distant (\textit{left}) and close (\textit{right}) contact binaries. The left diagram shows a distinct second population of highly reddened objects that cause the blue bump in the temperature distribution. Only objects with $\sigma_{T}<160$\,K and parallaxes with at least a $3\sigma$ precision are used.}
\label{bump_proof}
\end{figure}

\begin{figure}[!h]
\centering
\includegraphics[width=14cm]{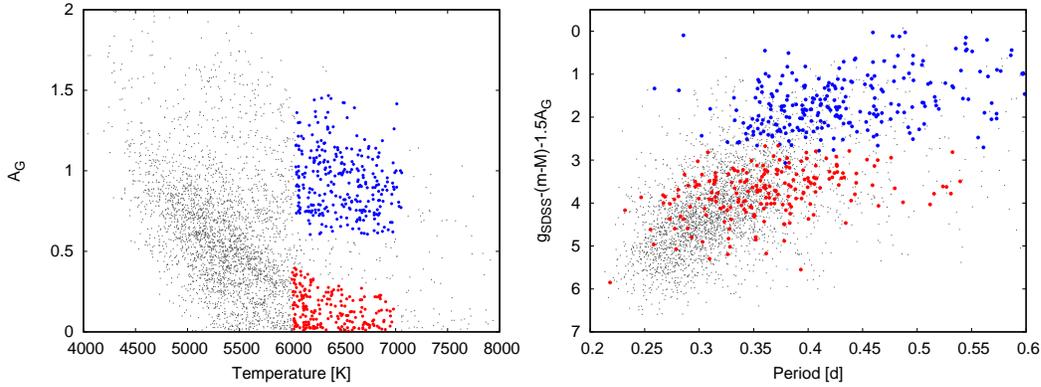}
\caption{\textit{Left:} Two subsections were made on the temperature-extinction diagram in the $6000-7000$\,K range: object with high and low extinction (blue and red circles, respectively). \textit{Right:} Orbital period vs. estimated Sloan $g$ absolute magnitude (using the rough approximation that $A_{g}=1.5A_{G}$). The objects with high exctinction, which cause the blue bump, have systematically larger luminosities as well as slightly larger periods. The distance moduli are calculated by inverting the Gaia parallaxes. Only objects with $\sigma_{T}<160$\,K and parallaxes with at least a $3\sigma$ precision are used.}
\label{bump_redblue}
\end{figure}

Instead, we argue that the blue bump in the temperature distribution is an artificial peak, caused by incorrect classification of other types of variables as contact binaries. We plotted SDSS temperatures vs. Gaia G-band extinction estimates for various distances (Fig.\,\ref{bump_proof}). The plots show a noisy main sequence complemented by a distinct second population of objects with a high extinction. These become prominent at larger distances, where the blue bump appears (Fig.\,\ref{tempdist}). These objects have systematically higher luminosities and slightly longer periods than contact binaries with the same temperatures at low extinctions (Fig.\,\ref{bump_redblue}). Our estimated g-band absolute magnitudes show that the majority of objects with a low extinction are in the $M_{g}=3-4$\,mag range, which is normal for F-type contact binaries. 

Noisy lightcurves of semi-detached binaries may be easy to confuse with W UMa-type lightcurves and semi-detached binaries are expected to be systematically hotter as it takes time for a close system to evolve into a contact stage. Therefore our main hypothesis is that objects with a higher extinction, which are the ones causing the blue bump in the temperature distribution, have been systematically misclassified as contact binaries. That would mean that the actual temperature distribution of contact binaries does not contain a bump in the $6000-7000$\,K range.   

\section{Summary}
\label{sect:discussion}
We combined data from the VSX, SDSS DR12, and Gaia DR2 catalogs to explore the observed temperature distribution of known contact binaries. We used color-temperature relations to calculate the effective temperatures of the matched set and built a temperature distribution for $\sim13000$ contact binaries. We also showed how the distribution shifts to higher temperatures when increasing the distance of the observed objects. The distribution contains a bump in the $6000-7000$\,K range previously detected by \citet{2017RAA....17...87Q}. We show that this second maximum is caused by distant and highly reddened objects that are systematically more luminous than contact binaries in this temperature range. We put forward the idea that these objects are other variables, possibly semi-detached binaries, misclassified as contact binaries and that the actual temperature distribution of contact binaries does not contain such a peak.

\normalem
\begin{acknowledgements}
This work was supported by grants DFNP-17-13 and DFNP-17-70 of the Young Scientists Support Program at the Bulgarian Academy of Sciences. We acknowledge the use of archive data from GAIA and are grateful to the GAIA team. Funding for SDSS-III has been provided by the Alfred P. Sloan Foundation, the Participating Institutions, the National Science Foundation, and the U.S. Department of Energy Office of Science. The SDSS-III web site is http://www.sdss3.org/. SDSS-III is managed by the Astrophysical Research Consortium for the Participating Institutions of the SDSS-III Collaboration including the University of Arizona, the Brazilian Participation Group, Brookhaven National Laboratory, Carnegie Mellon University, University of Florida, the French Participation Group, the German Participation Group, Harvard University, the Instituto de Astrofisica de Canarias, the Michigan State/Notre Dame/JINA Participation Group, Johns Hopkins University, Lawrence Berkeley National Laboratory, Max Planck Institute for Astrophysics, Max Planck Institute for Extraterrestrial Physics, New Mexico State University, New York University, Ohio State University, Pennsylvania State University, University of Portsmouth, Princeton University, the Spanish Participation Group, University of Tokyo, University of Utah, Vanderbilt University, University of Virginia, University of Washington, and Yale University.

\end{acknowledgements}
  
\bibliographystyle{raa}
\bibliography{WUMa_distrib}

\end{document}